# Comment: Expert Elicitation for Reliable System Design

**Wenbin Wang**

## INTRODUCTION

The paper by Bedford, Quigley and Walls has reviewed the role of expert judgements in reliability assessment within the engineering design process. It differs from many statistically based papers that focus mainly on statistical models, in that it addresses issues related to the process, the concept and principles. The paper itself is comprehensive in coverage and offers general guidance for reliability assessment at the system design stage. However, to some extent, the technical side of the statistical techniques used in expert opinion elicitation in reliability analysis has not been fully explored. A practitioner who wants to conduct research in the area still needs to search methodologies from other recommended sources in the paper to find practical solutions. This is perhaps the purpose of the paper.

In my discussion, two techniques that are particularly useful to subjective data elicitation, but are not discussed in the Bedford, Quigley and Walls paper, namely, empirical Bayes (EB; Robbins, 1955) and evidential reasoning (ER; Dempster, 1966), will be briefly discussed. It is hoped that this will shed some light on the problem of expert data elicitation in general and reliability assessment in design in particular.

## THE EB AND ER APPROACHES

Suppose that a prior distribution is hypothesized for the parameter $\Theta$, but is only specified to be in a certain class of prior distributions. We can represent this by a prior distribution $p(\Theta|\Phi)$, where the hyperparameters $\Phi$ index the family of priors. We can then construct a "likelihood"

$$p(x|\Phi) = \int p(x|\Theta)p(\Theta|\Phi)\,d\Theta \tag{1}$$

that relates the data to the hyperparameters. In the so-called empirical Bayes approach, $\Phi$ is estimated from (1) by classical methods such as unbiased estimation, yielding an estimate $\hat{\Phi}$. The prior is then taken to be $p(\Theta|\hat{\Phi})$ and the inference about $\Theta$ is by the appropriate Bayes rule for this prior. The analysis thus has the flavor of a Bayesian analysis, but with an empirical prior based on the data, so it is termed empirical Bayes. If we view $x$ as the subjective or expert data, then the EB approach can be readily used in the reliability analysis of design problems.

*Wenbin Wang is Senior Lecturer, Centre for Operational Research and Applied Statistics, University of Salford, Salford, M5 4WT, United Kingdom, and also is affiliated with the School of Management, Harbin Institute of Technology, Harbin, China e-mail: w.wang@salford.ac.uk.*

The Dempster–Shafer theory of evidence (Dempster, 1966; Shafer, 1976) was particularly developed to aggregate subjective assessments in decision-making. To begin with, suppose there is a simple two-level hierarchy of attributes with a general attribute at the top and a number of basic attributes at the bottom. In the reliability assessment of a designed system, the basic attributes can be viewed as the components of the system and the general attribute can be the system reliability. Let $P_{n,i}$ be a probability mass that represents the degree to which the $i$th basic attribute supports the hypothesis that the general attribute is assessed to the $n$th grade. Then $P_{n,i}$ is calculated as

$$P_{n,i} = \omega_i \beta_{n,i}, \tag{2}$$

where $\omega_i$ is the weight of the $i$th attribute and $\beta_{n,i}$ is the degree of belief that the $i$th basic attribute supports the hypothesis that the general attribute is assessed to the $n$th grade. Define a subset of the basic attributes and let $P_{n,I(i)}$ denote the probability mass defined as the degree to which all the attributes in the subset support the hypothesis that the general attribute is assessed to the $n$th grade. Then we have

$$P_{n,I(i+1)} = K_{I(i+1)}P_{n,I(i)}P_{n,i+1}, \tag{3}$$

where $K_{I(i+1)}$ is a normalizing factor. In the original ER approach, the combined degree of belief $\beta_n$ is given by

$$\beta_n = P_{n,I(L)},$$ (4)

where $L$ is the number of basic attributes.

For simplicity, (3) is presented by assuming $\sum_{i=1}^{N} P_{n,i} = 1$, where $N$ is the number of grades. The case of $\sum_{i=1}^{N} P_{n,i} < 1$ was actually established in the original ER algorithm (Shafer, 1976).

From this brief introduction to the EB and ER approaches, we can see that they are relevant to reliability assessment using subjective expert data. In the next section, we further discuss these two techniques with respect to the reliability assessment problem at the design stage.

## EB AND ER APPROACHES IN RELIABILITY ASSESSMENT

We start with the EB approach, which is mainly concerned with the Bayesian treatment of the data collected for reliability assessment.

It is noted that the most difficult problem in subjective data acquisition is formulating the type of questions designed to collect useful data from the experts. I agree with the authors that the best approach is to ask the experts information about their observed events, which is typically done in probabilistic risk assessment (PRA). Of course, at the design stage many events are unknown; therefore, we have to extract information from the experts about the unknown quantity. Empirical Bayes methods may be of use in this context; see Lemon (1972), Yüceer and Inoue (1980), Tillman, Kuo, Hwang and Grosh (1982), Savchuk (1988), Ho and Pinheiro (2002), Denson, Keene and Caroli (1998), Vasconcelos and Lippman (1997) and Wang and Jia (2006).

The EB technique uses information from observed events, but the aim is to make inference on the unobserved quantity of interest. In standard Bayesian methods, priors are typically chosen so as to minimize computational complexity or are set to arbitrary values. While using empirical Bayes methods, the hyperparameters in the prior are obtained by maximizing the marginal distribution of some observed events conditional on the hyperparameters; see (1). This approach may violate the fundamental Bayesian principle that priors should not be estimated from the data, but in practice it leads to more sensible solutions than setting priors arbitrarily or using priors whose main justification comes from computational simplicity (the so-called conjugate priors). Wang and Jia (2006) have demonstrated this point using simulated and real data. The empirical Bayes approach also provides a way to break the infinite chain of conditional probabilities involved in standard hierarchical Bayesian inference, while it still allows for different priors, depending on the context. The EB approach is an approximation to the standard Bayesian analysis, but it still retains some flavor of the Bayesian treatment as explained before. The main advantage of the EB approach over standard Bayesian analysis is the reduction of computing complexity and time requirements. This is particularly useful when handling a design problem with many unknown factors. The accuracy of such approximation depends on the case considered and the shapes of the prior distributions.

Consider the following example: If we are interested in design reliability $d$, which is characterized by an unknown parameter $\Theta$ and a prior distribution $p(\Theta|\Phi)$, where $\Phi$ are the hyperparameters of the prior, it follows that

$$p(d|\Phi) = \int p(d|\Theta)p(\Theta|\Phi)\,d\Theta.$$ (5)

Suppose that a set of expert data $x$ is available. Then, using Bayes' theorem, we have

$$p(\Theta|x) = \frac{p(x|\Theta)p(\Theta)}{\int p(x|\Theta)p(\Theta)\,d\Theta}.$$ (6)

Since $\Theta$ is unknown in a standard framework of Bayesian inference unless there is absolute certainty regarding the values of $x$, we have to use

$$p(\Theta|x) = \frac{\int p(x|\Theta)p(\Theta|\Phi)p(\Phi)\,d\Phi}{\int \int p(x|\Theta)p(\Theta|\Phi)p(\Phi)\,dx\,d\Phi}$$ (7)

instead of (6). This will, of course, involve a significant increase in complexity.

The empirical Bayes perspective is to avoid this increase by keeping $p(\Theta|\Phi)$ but choosing the parameter set $\Phi$ that best explains the data $x$ from (1). Inferences are then based on (6) using these estimated values, namely,

$$p(\Theta|x, \hat{\Phi}) = \frac{p(x|\Theta)p(\Theta|\hat{\Phi})}{\int p(x|\Theta)p(\Theta|\hat{\Phi})\,d\Theta}.$$ (8)

Equation (5) then becomes

$$p(d|\hat{\Phi}) = \int p(d|\Theta)p(\Theta|x, \hat{\Phi})\,d\Theta.$$ (9)

Now we discuss the potential use of ER in reliability assessment at the design stage. The ER approach is particularly useful in reliability assessment when the data collected are imprecise and incomplete, which is often encountered in subjective data elicitation.

Data obtained from experts are typically imprecise and incomplete. This is a key issue in subjective data elicitation. This topic is hot, particularly in the area of decision-making using subjective data; see, for example, Athanassopoulos and Podinovski (1997), Bana e Costa and Vincke (1995), Miettinen and Salminen (1999) and Weber (1987). Coolen and Newby (1994) also provides a good treatment of the same issue in reliability analysis from a Bayesian point of view. Imprecise information means that the information provided does not impose a precise combination of the values for the parameters of interest. A new designation for this type of information is the name nonpoint information, which implies that multiple scenarios and interval data are typically the type of data collected. A good deal of research carried out in multi-criteria decision-making using evidential reasoning (ER) could be of use in this context of reliability assessment; see Yang and Xu (2002), Palacharla and Nelson (1994), Saffiotti (1994), Dempster (1966), Dempster and Kong (1986) and Shafer (1976). One feature of ER is its ability to handle qualitative data, which can also be useful in reliability assessment. Suppose for now that the new system under design is a motorcycle. To evaluate the reliability of this system, the general attribute can be the reliability of the motorcycle, and the three basic attributes are assumed to be engine, transmission and brakes. The set of evaluation grades is defined by H = {poor, indifferent, average, good, excellent}. Reliability is a general technical concept and it is difficult to assess directly. It needs to be decomposed into detailed subconcepts. Even the concept of brakes must be measured by stopping power, braking stability and feel at control. Now if expert data become available with respect to an expert's personal degree of belief, $\beta_{n,i}$, that the $i$th basic attribute supports the hypothesis that the general attribute is assessed to the $n$th grade and the weight $\omega_i$ is also available, then by utilizing (2)–(4), the reliability of the new motorcycle can be assessed. There is a considerable amount of literature on the selection of appropriate weights; see Hwang and Yoon (1981), Saaty (1988) and Yang, Deng and Xu (2001).

Reasoning about uncertainty emphasizes the practical application of theoretically sound techniques for reasoning from evidence based on information that is potentially incomplete, inexact, inaccurate and from diverse sources. Although there is very little reported research on the application of ER to reliability assessment, the ER technique has been widely used in Bayesian belief networks (as mentioned in the paper by Bedford, Quigley and Walls). In fact, the probabilistic theory behind ER is Bayesian, but ER also uses many techniques other than probabilities, such as possibility and fuzziness. The probability structure employed in ER is basically nonparametric and if a parametric form is used, EB can also be used.

## CONCLUSION

This discussion provides additional insight with regard to expert data elicitation in general, and applications to reliability assessment at the design stage in particular. The Bedford, Quigley and Walls paper covers the subject in a broad way, following a top-down approach. It outlines the general procedure, structure and issues related to the concepts in expert elicitation for reliability system design. It is a useful addition to the existing literature, but, as stated before, a detailed description of the statistical techniques used in expert elicitation can also be important.